\documentclass{pj}
\usepackage{graphicx}
\usepackage{amsmath,amssymb}

\newcommand{\ie}{\textit{i}.\textit{e}.}

\begin{document}
\setcounter{page}{1}
\pjheader{July 2, 2020}

\title[IPO: Iterative Physical Optics Image Approximation]
{IPO: Iterative Physical Optics Image Approximation}  \footnote{\hskip-0.12in*\, Corresponding
author:~Shaolin~Liao~ (sliao5@iit.edu).} 
\footnote{\hskip-0.12in\textsuperscript{1} S. Liao is with Department of Electrical and Computer Engineering, Illinois Institute of Technology, Chicago, IL 60616 USA. \textsuperscript{2} L. Ou (oulu9676@gmail.com) is with College of Computer Science and Electronic Engineering, Hunan University, Changsha, Hunan, China  410082.}

\author{Shaolin~Liao\textsuperscript{*, 1} and Lu~Ou\textsuperscript{2}}

\runningauthor{Liao and Ou}


\begin{abstract}
An improved Iterative Physical Optics (IPO) image approximation method has been presented to dramatically increase the accuracy of the approximation and extend its applicability to PEC surfaces with smaller radii or larger curvatures. Starting from the first-order conventional PO image approximation, the IPO image approximation method iteratively correct the surface current to compensate the deviation of the electric field boundary condition on the PEC surfaces, making use of the local plane wave approximation. Numerical validations with two popular PEC surfaces, \ie, the parabolic dish antennas and the PEC spheres, are carried out and the results show that the IPO approximation method increases the surface current accuracy by more than two orders of magnitude, compared to the conventional PO image approximation method. 
\end{abstract}


\setlength {\abovedisplayskip} {6pt plus 3.0pt minus 4.0pt}
\setlength {\belowdisplayskip} {6pt plus 3.0pt minus 4.0pt}

\

\section{Introduction}
\label{sec:intro}   
Physical Optics (PO) image approximation has been widely used for smooth Perfect Electric Conductor (PEC) surfaces, due to its efficiency and accurate approximation for relatively smooth PEC surfaces \cite{liao_image_2006, liao_near-field_2006, shaolin_liao_new_2005, liao_fast_2006, liao_cylindrical_2006, liao_beam-shaping_2007, liao_fast_2007, liao_validity_2007, liao_high-efficiency_2008, liao_four-frequency_2009, vernon_high-power_2015, liao_multi-frequency_2008, liao_fast_2007-1, liao_sub-thz_2007, liao_miter_2009, liao_fast_2009, liao_efficient_2011, liao_spectral-domain_2019}. Compared to other advanced Computational Electromagnetics (CEM) methods such as the Method of Moments (MoM), the application of the PO image approximation can dramatically reduce the the number of unknowns and memory requirements for the electromagnetics scattering problem of electrically large objects \cite{ma_efficient_2012}. The PO image approximation can find important applications in antennas design and analysis \cite{chuan_liu_design_2013, yurduseven_compact_2011},  beam-shaping \cite{liao_beam-shaping_2007, liao_high-efficiency_2008, liao_four-frequency_2009, vernon_high-power_2015, liao_multi-frequency_2008, liao_fast_2007-1, liao_sub-thz_2007, liao_miter_2009},  and Radar Cross Section (RCS) calculation \cite{emhemmed_analysis_2019, wang_radar_2017}. In particular, the author's group has applied the PO method to design a beam-shaping mirrors system  at the millimeter-wave regime to shape the multi-mode TE waves (TE$_{22,6}$/110 GHz,  TE$_{23,6}$/113 GHz,  TE$_24,6$/116 GHz  and  TE$_25,6$/118.8 GHz) into the Gaussian beams, resulting in efficiencies $> 98\%$ \cite{liao_beam-shaping_2007, liao_high-efficiency_2008, liao_four-frequency_2009, vernon_high-power_2015, liao_multi-frequency_2008}. 

Although useful in the above important applications, the PO image method is only exact for planar PEC surface and is approximate for non-planar but relatively smooth PEC surfaces: the larger radii or the smaller the curvatures of the surfaces, the better the approximation \cite{liao_image_2006, liao_near-field_2006}. So it would be beneficial if the PO image approximation can be extended to PEC surfaces of smaller radii or larger curvatures with satisfactory accuracy, which is the focus of this letter.

\begin{figure}[th]
 \centering
 \includegraphics[width=0.8\textwidth]{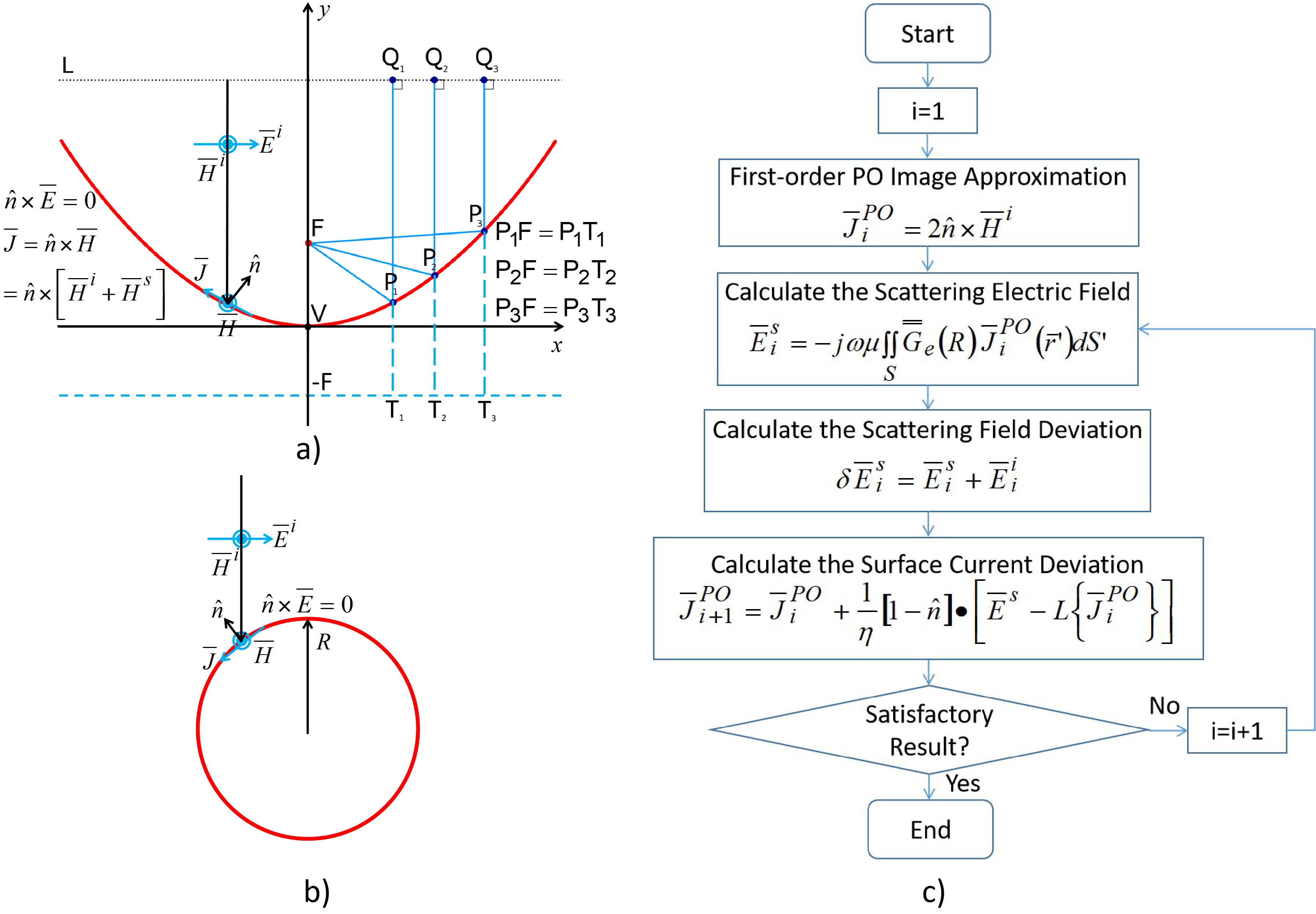}
\caption{The improved IPO approximation method for the electromagnetic scattering from relatively smooth PEC surfaces: a) the parabolic dish antenna; b) the PEC spheres;; and c) the IPO algorithm.}
\label{fig:problem}
\end{figure}

\section{Problem Formulation}\label{sec:problem}
The electromagnetic scattering problem from a PEC surface is shown in Fig. In general, the unknown surface current can only be solved through rigorous CEM methods such as MoM by imposing zero total electric field boundary condition on the metallic surface as follows,
\begin{flalign}\label{eqn:BC}
\hat{n} \times \left[\overline{E}^i (x, y) +  \overline{E}^s (x, y) \right]  =0,
\end{flalign}
where $\hat{n}$ is the unit surface normal of the PEC surface; also, the superscripts $s$ and $i$ denote the scattering and incident electric field, respectively. 

For perfect planar surface, the exact surface current can be obtained by the PO as follows,
\begin{flalign}\label{eqn:PO}
\overline{J}^{PO} = \hat{n} \times \overline{H} = \hat{n} \times \left\{ \overline{H}^{i}  + \overline{H}^{s} \right\} =  2\hat{n} \times \overline{H}^{i},
\end{flalign}
and the boundary condition in Eq. (\ref{eqn:EFIE}) reduces to the following,
\begin{flalign}\label{eqn:BC2}
\hat{n} \times \left[\overline{E}^i (x, y) +  \mathcal{L} \left\{ \overline{J}^{PO} \right\} \right]   = \hat{n} \times  \left[\overline{E}^i (x, y) +  \mathcal{L} \left\{   2\hat{n} \times \overline{H}^{i} \right\} \right]  = 0,
\end{flalign}
where $\mathcal{L}$ is the operator that computes the scattering electric field form the surface current.

For smooth PEC surfaces, PO is only approximate \cite{liao_image_2006, liao_near-field_2006, liao_multi-frequency_2008}. So it would be beneficial if the accuracy of the conventional PO image approximation method can be improved so that it can be extended to PEC surfaces with smaller radii or larger curvatures.

\section{The Scattering Electric Field}
 The scattering electric field $\overline{E}^s$ can be expressed in terms of convolution of the surface current $\overline{J}$ and the electric dyadic Green's function $\overline{\overline{G}}_e$ as follows,
\begin{flalign}\label{eqn:field_convolution}
\overline{E}^s  =  \mathcal{L} \left\{  \overline{J} \right\} =  \overline{\overline{G}}_e \circledast \overline{J}  = - j \omega \mu \iint\limits_{S} \overline{\overline{G}}_e (R) \overline{J}(\overline{r}') dS',  
\end{flalign}
where $\circledast$ denotes the 2D convolution operation; $R = r- r'; r=|\overline{r}|, r'=|\overline{r}'|$ with $\overline{r}$ and $\overline{r}'$ being the observation point and source point respectively; $\omega$ is the angular frequency; $\mu$ is the permeability; and the dyadic Green's function is given as Eq (\ref{eqn:dyadic_green}),
\begin{flalign}\label{eqn:dyadic_green}
 & \overline{\overline{G}}(\overline{r}) =   g(\overline{r}) \overline{\overline{I}} + \frac{1}{\left(k\right)^2} \nabla \nabla  g(\overline{r}); \ \  g(r) = \frac{e^{-j k r }}{4 \pi r}, \ \ r = \left|\overline{r}\right|; \ \ k =|\overline{k}| = \omega \sqrt{\mu \epsilon},
\end{flalign}
with $\overline{\overline{I}}$ being the identity matrix and $k$ being the magnitude of the wave vector $\overline{k} = [k_x, k_y, k_z]$.

\section{The EFIE}
With the scattering electric field given in Eq. (\ref{eqn:field_convolution}), the EFIE equation can be obtained from the boundary condition of Eq. (\ref{eqn:BC}) as follows, 
\begin{flalign}\label{eqn:EFIE}
\hat{n} \times  \left[\overline{E}^i (x, y) - j \omega \mu \iint\limits_{S} \overline{\overline{G}}_e (R) \overline{J}(\overline{r}') dS' \right]  = 0,
\end{flalign}
where the surface current $\overline{J}$ remains to be solved.

\section{The Iterative PO (IPO) Approximation}
The first-order IPO approximation for the surface current $\overline{J}$ is the PO image theorem given in Eq. (\ref{eqn:PO}),
\begin{flalign}\label{eqn:PO1}
\overline{J}_1^{IPO} =  \overline{J}^{PO} =  2\hat{n} \times \overline{H}^{i}.
\end{flalign}

Now the deviations of the boundary condition of Eq. (\ref{eqn:BC}) is given by,
\begin{flalign}\label{eqn:dE}
\hat{n} \times \delta \overline{E}_1^s = \hat{n} \times \left[ \overline{E}^s - \mathcal{L} \left\{ \overline{J}_1^{IPO}  \right\} \right].
\end{flalign}

Approximating the local electromagnetic field as local plane wave, the electric field $ \delta \overline{E}_1^s$ is related to the deviation of the deviation of the magnetic field $ \delta \overline{H}_1^s$ as follows,
\begin{flalign}\label{eqn:dH}
 \delta \overline{E}_1^s     =   \eta  \delta \overline{H}_1^s  \times \hat{n}; \ \ \eta = \sqrt{\mu/\epsilon}.
\end{flalign}

Substituting Eq. (\ref{eqn:dH}) into Eq. (\ref{eqn:dE}), the following is obtained,
\begin{flalign}\label{eqn:dEH}
\hat{n} \times   \left[ \delta \overline{H}_1^s  \times \hat{n} \right]  = \frac{1}{\eta} \hat{n} \times \left[ \overline{E}^s - \mathcal{L} \left\{ \overline{J}_1^{IPO}  \right\} \right] \rightarrow   \delta \overline{H}_1^s   = \frac{1}{\eta} \hat{n} \times \left[ \overline{E}^s - \mathcal{L} \left\{ \overline{J}_1^{IPO}  \right\} \right].
\end{flalign}

Now, to compensate the deviations of the electric field $\delta \overline{E}_1^s$, the surface current can be corrected   $\delta \overline{J}_1^{IPO}$ as follows, 
\begin{flalign}
\delta \overline{J}_1^{IPO} = - \hat{n} \times  \delta \overline{H}_1^s =  - \frac{1}{\eta} \hat{n} \times \left\{ \hat{n} \times \left[ \overline{E}^s - \mathcal{L} \left\{ \overline{J}_1^{IPO}  \right\} \right] \right\},
\end{flalign}
from which  the deviations of the electric field $\delta \overline{E}_1^s$ is obtained as follows,
\begin{flalign}\label{eqn:current_iteration}
\delta \overline{J}_1^{IPO} = \frac{1}{\eta} \delta \overline{E}_{1}^{s//} =  \frac{1}{\eta}  \left[ 1- \hat{n} \right]  \bullet \left[ \overline{E}^s - \mathcal{L} \left\{ \overline{J}_1^{IPO}  \right\} \right],
\end{flalign}
and $\delta \overline{E}_{1}^{s//}$ is the tangential electric field projection on the PEC surface.
 
The surface current correction of Eq. (\ref{eqn:current_iteration}) is done iteratively until satisfactory result is obtained,
\begin{flalign}\label{eqn:current_update}
\delta \overline{J}_i  = \frac{1}{\eta} \left[ 1- \hat{n} \right]  \bullet \left[ \overline{E}^s - \mathcal{L} \left\{ \overline{J}_i^{IPO}  \right\} \right].
\end{flalign}


\begin{figure}[th]
 \centering
 \includegraphics[width=1\textwidth]{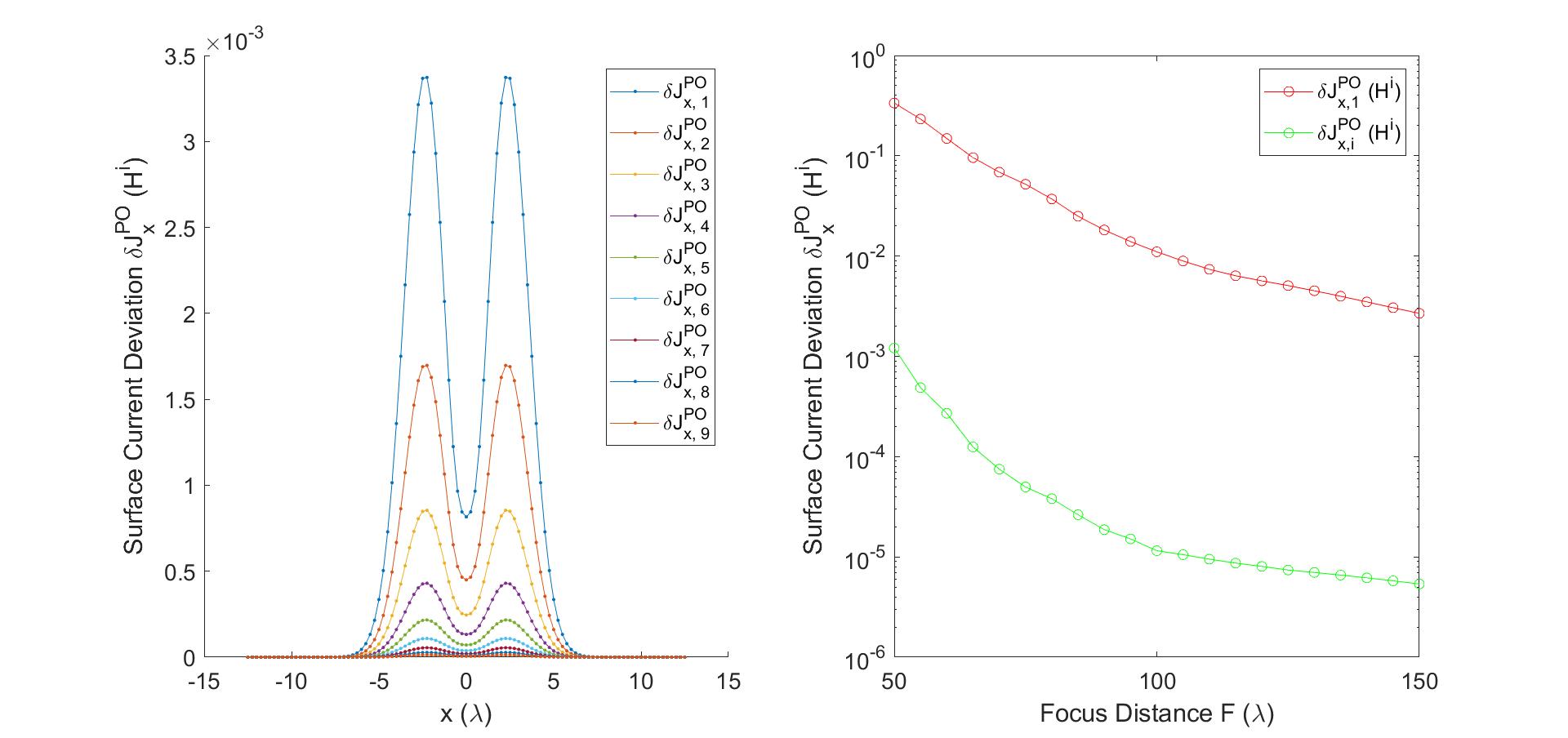}
\caption{Parabolic dish antennas: left) Surface current deviation $\eta \left| \delta \overline{J}_{x, i}^{IPO} \right|$ at different iterations of the IPO approximation method for the focus length of $F=100 \lambda$; and right) Surface current deviation of the conventional PO image approximation method $\eta \left| \delta \overline{J}_x^{PO} \right|$  and the IPO approximation method $\eta \left| \delta \overline{J}_x^{IPO} \right|$ for parabolic dish antennas of different focus lengths $F$.}
\label{fig:parabolic}
\end{figure}

\section{Algorithm}
Fig. \ref{fig:problem}c) shows the algorithm of the improved IPO image approximation: it starts with the first-order PO image approximation; then it calculates the scattering electric field according to Eq. (\ref{eqn:field_convolution}), followed by updating of the surface current according to Eq. (\ref{eqn:current_iteration}) and Eq. (\ref{eqn:current_update}); finally the algorithm ends when satisfactory result is met.

\begin{figure}[th]
 \centering
 \includegraphics[width=1.05\textwidth]{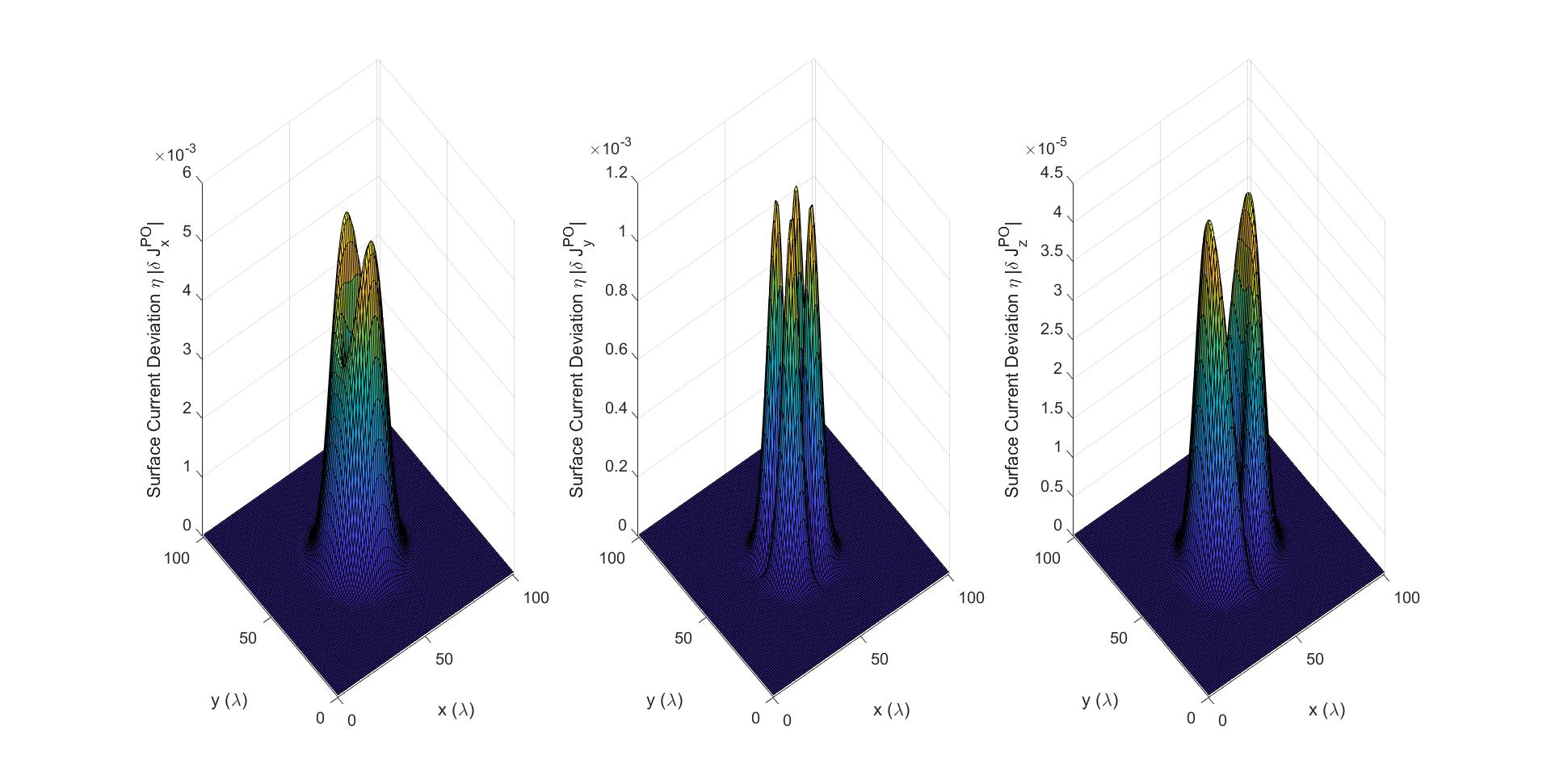}
\caption{PO surface currents deviation for a parabolic dish antenna  of a  focus length of $F = 100 \lambda$ (left to right): $\eta \left| \delta \overline{J}_x^{PO} \right|$; $\eta \left| \delta \overline{J}_y^{PO} \right|$; and $\eta \left| \delta \overline{J}_z^{PO} \right|$.}
\label{fig:dJ_PO}
\end{figure}

\begin{figure}[th]
 \centering
 \includegraphics[width=1.05\textwidth]{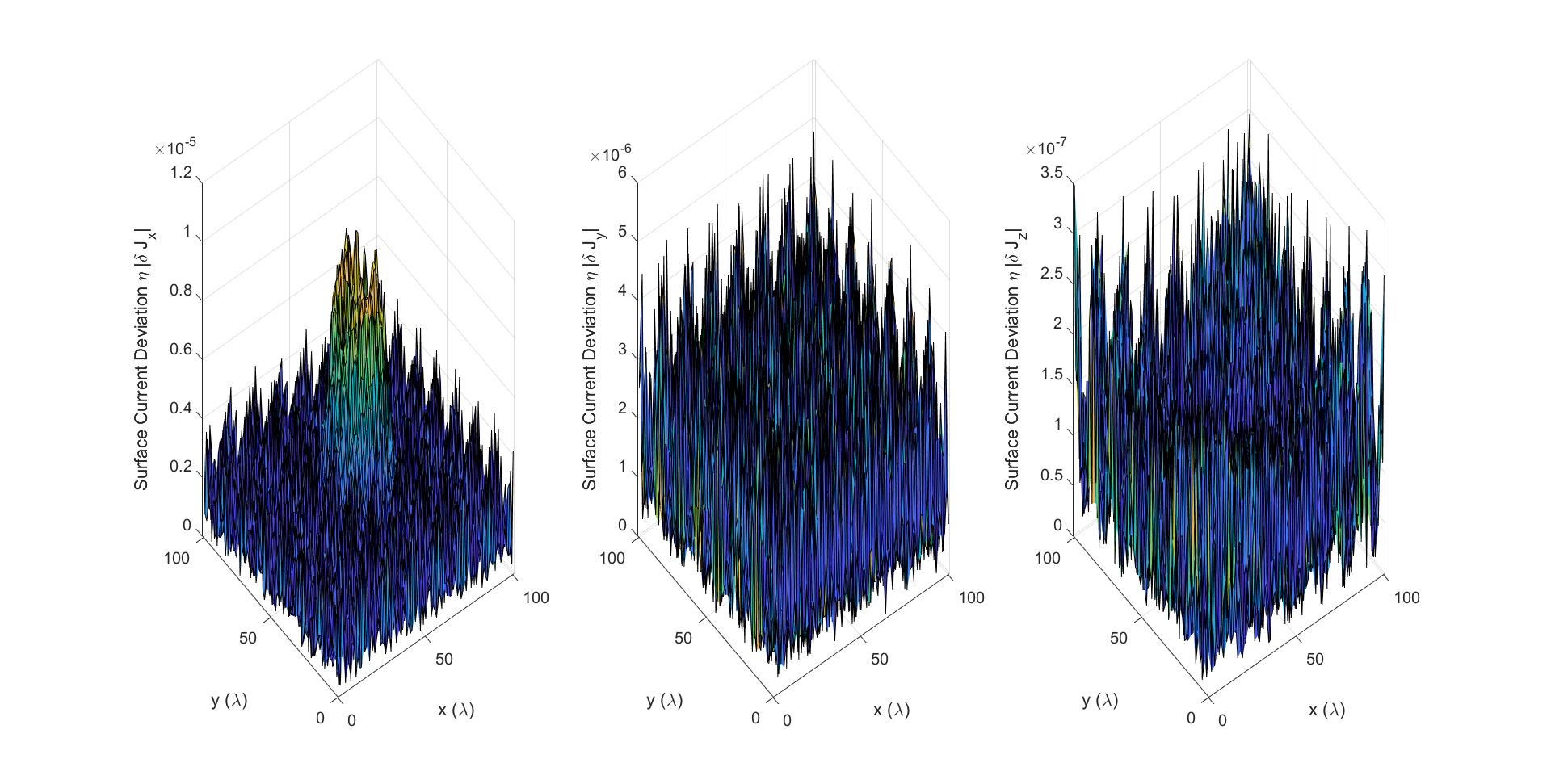}
\caption{IPO surface currents deviation for a parabolic dish antenna  of a  focus length of $F = 100 \lambda$ (left to right): $\eta \left| \delta \overline{J}_x^{PO} \right|$; $\eta \left| \delta \overline{J}_y^{PO} \right|$; and $\eta \left| \delta \overline{J}_z^{PO} \right|$.}
\label{fig:dJ_IPO}
\end{figure}

\begin{figure}[th]
 \centering
 \includegraphics[width=1\textwidth]{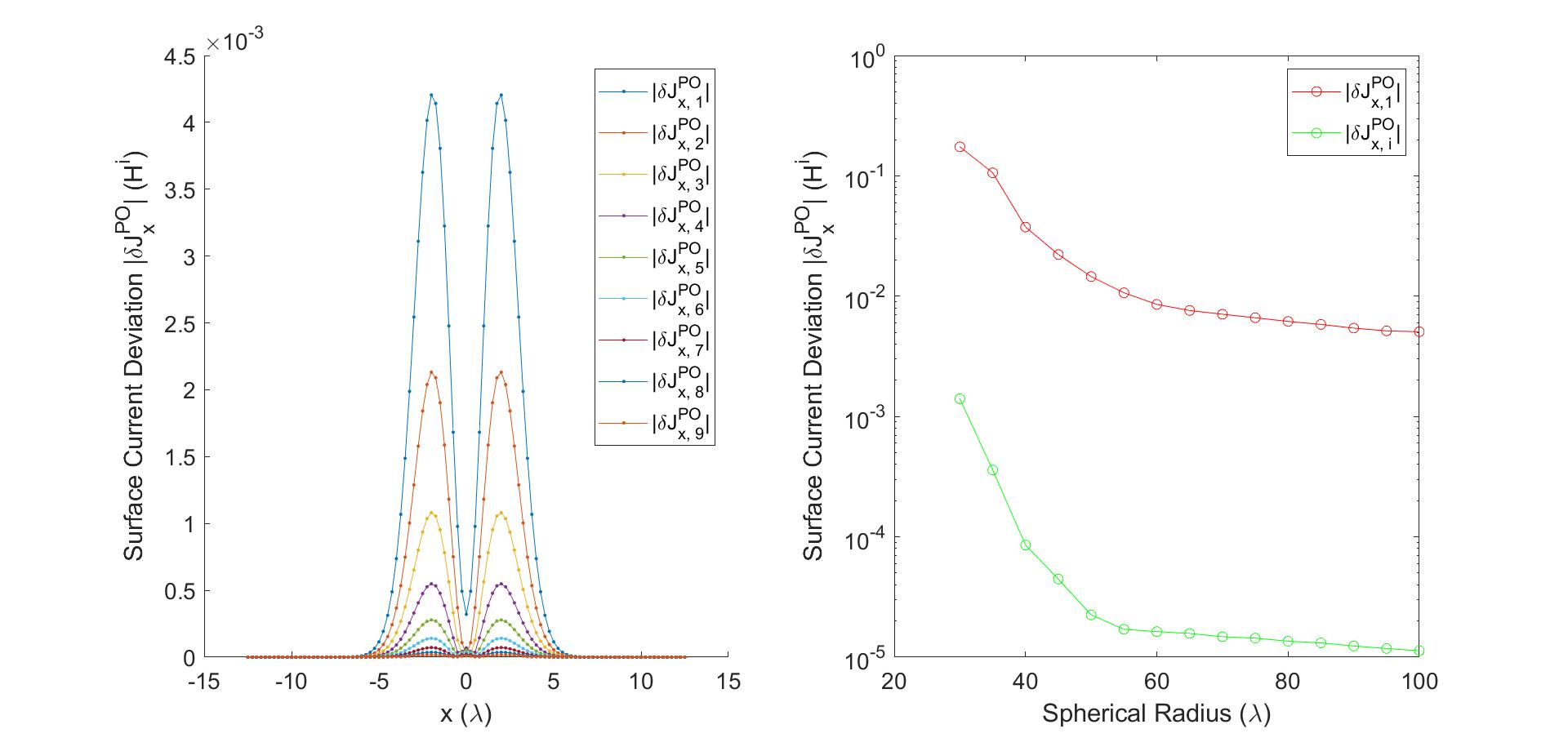}
\caption{PEC spheres: left) Surface current deviation $\eta \left| \delta \overline{J}_{x, i}^{IPO} \right|$ at different iterations of the IPO approximation method for the radius of $R = 60 \lambda$; and right) Surface current deviation of the conventional PO image approximation method $\eta \left| \delta \overline{J}_x^{PO} \right|$  and the IPO approximation method $\eta \left| \delta \overline{J}_x^{IPO} \right|$ for PEC spheres of different focus radii $R$.}
\label{fig:sphere}
\end{figure}

\begin{figure}[th]
 \centering
 \includegraphics[width=1.05\textwidth]{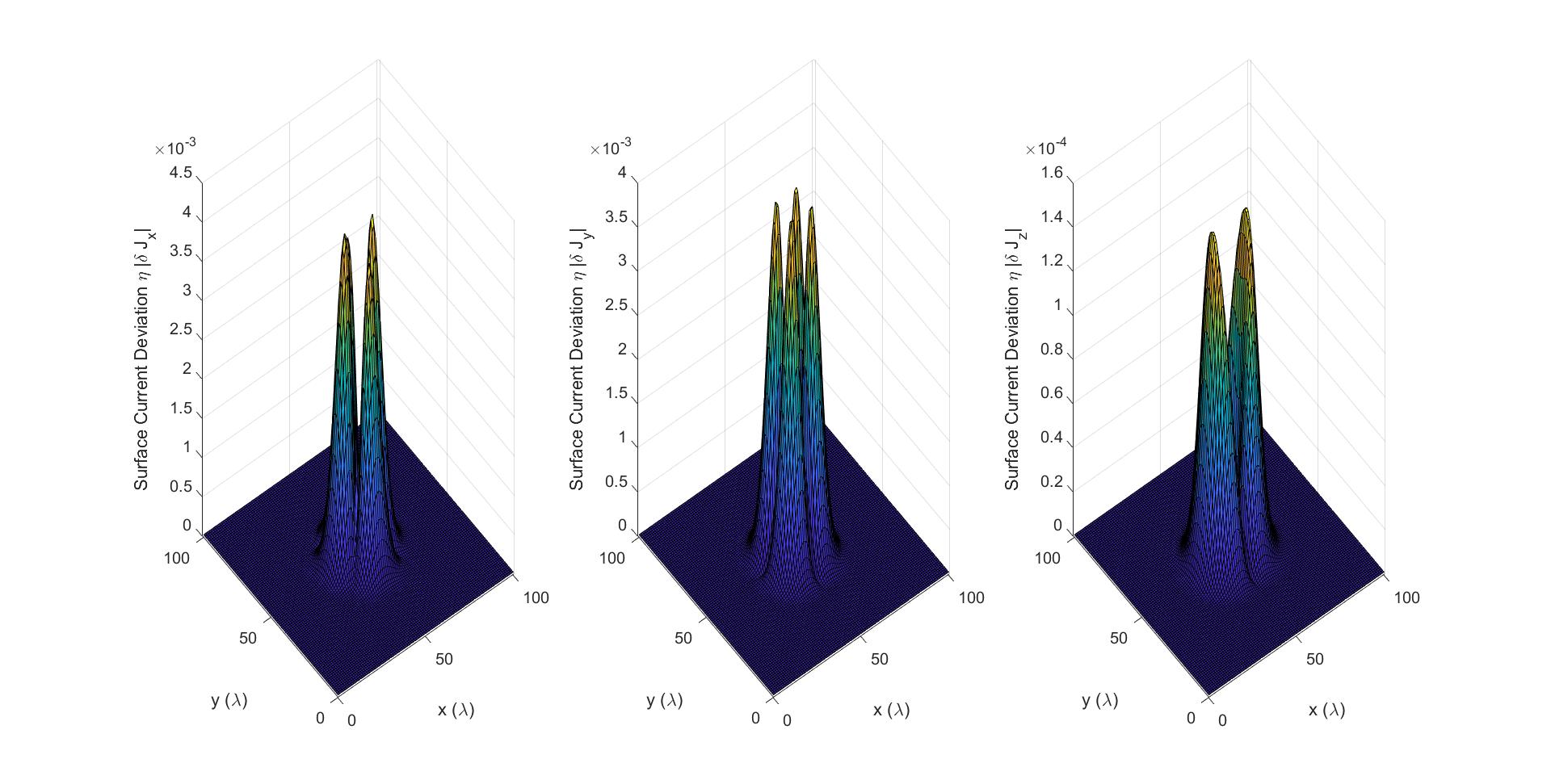}
\caption{PO surface currents deviation for a PEC sphere  of a  radius of $F = 60 \lambda$ (left to right): $\eta \left| \delta \overline{J}_x^{PO} \right|$; $\eta \left| \delta \overline{J}_y^{PO} \right|$; and $\eta \left| \delta \overline{J}_z^{PO} \right|$.}
\label{fig:dJ_PO_sphere}
\end{figure}

\begin{figure}[th]
 \centering
 \includegraphics[width=1.05\textwidth]{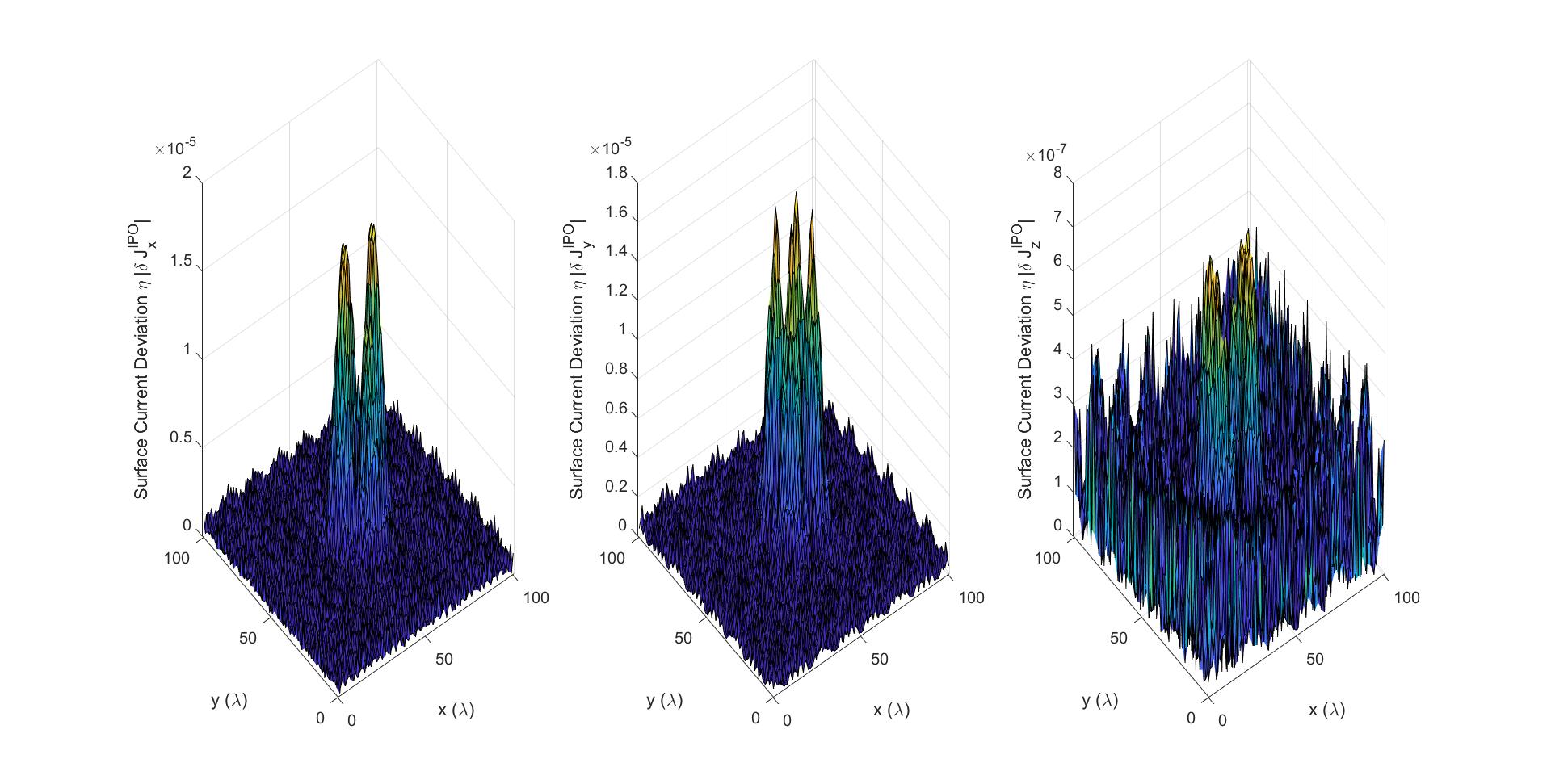}
\caption{IPO surface currents deviation for a PEC sphere  of a  radius of $F = 60 \lambda$  (left to right): $\eta \left| \delta \overline{J}_x^{PO} \right|$; $\eta \left| \delta \overline{J}_y^{PO} \right|$; and $\eta \left| \delta \overline{J}_z^{PO} \right|$.}
\label{fig:dJ_IPO_sphere}
\end{figure}

\section{Numerical Validation}
Two PEC surfaces are used to show the efficiency of the IPO image approximation with a 
Gaussian beam of waist $w = 2 \lambda$ as the incidence wave: 1) the parabolic dish antennas with different focus lengths $F$ (Fig. \ref{fig:problem} a); and 2) the PEC spheres of different radii $R$ (Fig. \ref{fig:problem}b). The PEC surface of the parabolic dish antenna \cite{chuan_liu_design_2013, yurduseven_compact_2011} and that of the PEC sphere are
\begin{flalign}
Z_{parabolic} = \frac{X^2 + Y^2}{ 4F}; \ \ Z_{spherical} = \sqrt{R^2 - X^2 -Y^2},
\end{flalign}
where $F$ is the focus distance of the parabolic dish antenna and $R$ is the radius of the PEC sphere.

The left plot of Fig. \ref{fig:parabolic} shows the surface current deviation $\eta \left| \delta \overline{J}_{x, i}^{IPO} \right|$ from the exact surface current obtained by the MoM at different iterations of the IPO approximation method for the parabolic dish antennas with a focus length of $F = 100 \lambda$, showing the convergence of the IPO method. Also, the right plot of Fig. \ref{fig:parabolic} shows the surface current deviations of the conventional PO image approximation method $\eta \left| \delta \overline{J}_{x}^{PO} \right|$ and that of those of the IPO approximation method $\eta \left| \delta \overline{J}_{x}^{IPO} \right|$ for different focus lengths $F = [50, 150] \lambda$, from which it can be seen that more than two orders of magnitude increase in accuracy has been achieved.

In addition, the surface current deviation of the PO image approximation and that of the IPO approximation for the parabolic dish antenna with a focus length of $F = 100 \lambda$ have been shown in Fig. \ref{fig:dJ_PO} and Fig. \ref{fig:dJ_IPO} respectively.

Similarly, the left plot of Fig. \ref{fig:sphere} shows the surface current deviation $\eta \left| \delta \overline{J}_{x, i}^{IPO} \right|$ at different iterations of the IPO approximation method for the PEC spheres with a radius of $R = 60 \lambda$; and the right plot of Fig. \ref{fig:sphere} shows the surface current deviations of the conventional PO image approximation method $\eta \left| \delta \overline{J}_x^{PO} \right|$ and those of the IPO approximation method $\eta \left| \delta \overline{J}_x^{IPO} \right|$ for different radii $R = [30, 100] \lambda$, from which it can be seen that the accuracy has been improved by more than two orders of magnitude also.

Finally, the surface current deviation of the PO image approximation and that of the IPO approximation for the PEC sphere with a radius of $R = 60 \lambda$ have been shown in Fig. \ref{fig:dJ_PO_sphere} and Fig. \ref{fig:dJ_IPO_sphere} respectively.

\section{Conclusion}\label{sec:con}
The improved IPO image approximation method has been presented to increase the accuracy of the conventional PO image approximation method. The IPO method iteratively corrects the surface current to compensate for the deviation of the electric field boundary condition on the PEC surfaces, assuming local plane wave approximation. Numerical experiments with parabolic dish antennas and PEC spheres show that the IPO method can increase the accuracy of the surface current by more than two orders of magnitude, compared to the conventional PO image approximation method.


\end{document}